\documentclass[
  twoside,
  twocolumn,
  aps,
  prl,
  preprintnumbers,
  floatfix,
  showpacs,
  citeautoscript,
]{revtex4-1}

\usepackage{amsmath}
\usepackage{amssymb}
\usepackage{graphicx}
\usepackage{times}
\usepackage{units}
\usepackage{glossaries}

\newcommand{\fig}[1]{Fig.~\ref{#1}}
\newcommand{\bccomg}{\text{$\text{\gls{bcc}}\rightarrow\omega$}}

\newacronym{bcc}{BCC}{body-centered cubic}
\newacronym{fcc}{FCC}{face-centered cubic}
\newacronym{hcp}{HCP}{hexagonal close-packed}
\newacronym{md}{MD}{molecular dynamics}
\newacronym{pes}{PES}{potential energy surface}

\begin{document}

\title{
  Defects from phonons: \\
  Atomic transport by concerted motion in simple crystalline metals
}

\author{Erik Fransson}
\author{Paul Erhart}
\email{erhart@chalmers.se}
\affiliation{
  Chalmers University of Technology,
  Department of Physics,
  S-412 96 Gothenburg, Sweden
}

\begin{abstract}
Point defects play a crucial role in crystalline materials as they do not only impact the thermodynamic properties but are also central to kinetic processes.
While they are necessary in thermodynamic equilibrium spontaneous defect formation in the bulk is normally considered highly improbable except for temperatures close to the melting point.
Here, we demonstrate by means of atomistic simulations that processes involving concerted atomic motion that give rise to defect formation are in fact frequent in body-centered cubic metals even down to about 50\%\ of the melting temperature.
It is shown that this behavior is intimately related to the anharmonicity of the lattice vibrations and a flat energy landscape along certain crystallographic directions, a feature that is absent in, e.g., face-centered cubic lattice structures.
This insight has implications for our general understanding of these materials and furthermore provides a complementary explanation for the so-called anomalous diffusion in group 4 transition metals.
\end{abstract}

\maketitle


Any real crystalline material contains defects, which are fundamental to its performance.
Specifically, point defects such as vacancies and interstitials are crucial for many transport properties including e.g., atomic diffusion and electrical conductivity.
Unlike one or two-dimensional defects such as dislocations or stacking faults, these zero-dimensional defects are thermodynamically necessary.
Especially in densely packed crystal structures such as \gls{fcc} or \gls{bcc} lattices, point defect concentrations are commonly established by exchange with an external reservoir such as a surface.
Defect formation can also be induced in the bulk by external means such as irradiation with ions or high energy photons \cite{AveDia98}.
By contrast, \emph{spontaneous} defect formation in the bulk is usually considered only close to the melting point \cite{SamTucYu14, DoaAdd87a}.

Theoretical studies of superheated crystals have revealed that collective atomic motion, driven by point defects, is critical for destabilizing the lattice and nucleate melting \cite{ZhaKhaLiu13, SamTucYu14}.
Studies of atomic dynamics in glasses and liquids have identified related processes where atoms move collectively in string-like fashion with neighboring atoms being unaffected \cite{NorAshAve05}.
In these systems this collective motion has been connected to the phonon dispersion as well as a pronounced boson peak \cite{ZhaKhaLiu13, ChuSerBur04, ShiTan08}.
Interestingly, phonons are also known to be related to self-diffusion in crystalline materials and, in particular, many studies have been carried out concerning the phonon driven diffusion in \gls{bcc} metals \cite{Her93, HerKoh87_2, KoeHer87, HerKOh87, VogPetFlo89, HerKohDiv99}.

Here, we demonstrate that spontaneous defect formation occurs in \gls{bcc} metals with high frequency down to about 50\%\ of the melting temperature.
Our analysis reveals that the nucleation of vacancy-interstitial (Frenkel) pairs is the result of concerted atomic motion along close-packed $\left<111\right>$ directions, similar to the collective atomic motion in superheated crystals, glasses and liquids.
We establish a clear connection between these processes and specific phonon modes that turn out to be the same phonon modes that have previously been analyzed in connection with diffusion \cite{Her93, HerKoh87_2, KoeHer87, HerKOh87, VogPetFlo89, PetFloHei88}.
This enables us to identify the key features of the \gls{pes}, which are responsible for the observed behavior, and reveal trends across groups of materials.
The insight obtained in this fashion not only has consequences for our general understanding of defects but suggests a complementary explanation for the non-Arrhenius-like self-diffusion in group 4 metals.
Previous studies have suggested that the strong temperature dependency of phonons gives rise to a temperature dependent migration of defects \cite{Her93, HerKoh87_2, KoeHer87, MikOse93}.
The present results, however, indicate that the defect concentration itself (and not just the migration rate) can strongly deviate from an exponential temperature dependence.

In the following, we first describe the variation of defect pair concentrations as a function of temperature for a range of metals.
Then, we resolve the atomistic mechanisms that gives rise to spontaneous defect formation and establish a relation to phonon dispersion and \gls{pes}.
Finally, we discuss the implications of the results and the effect of \gls{pes} and lattice structure.
Computational details can be found in the Appendix.


\begin{figure}
  \centering
  \includegraphics[scale=0.5]{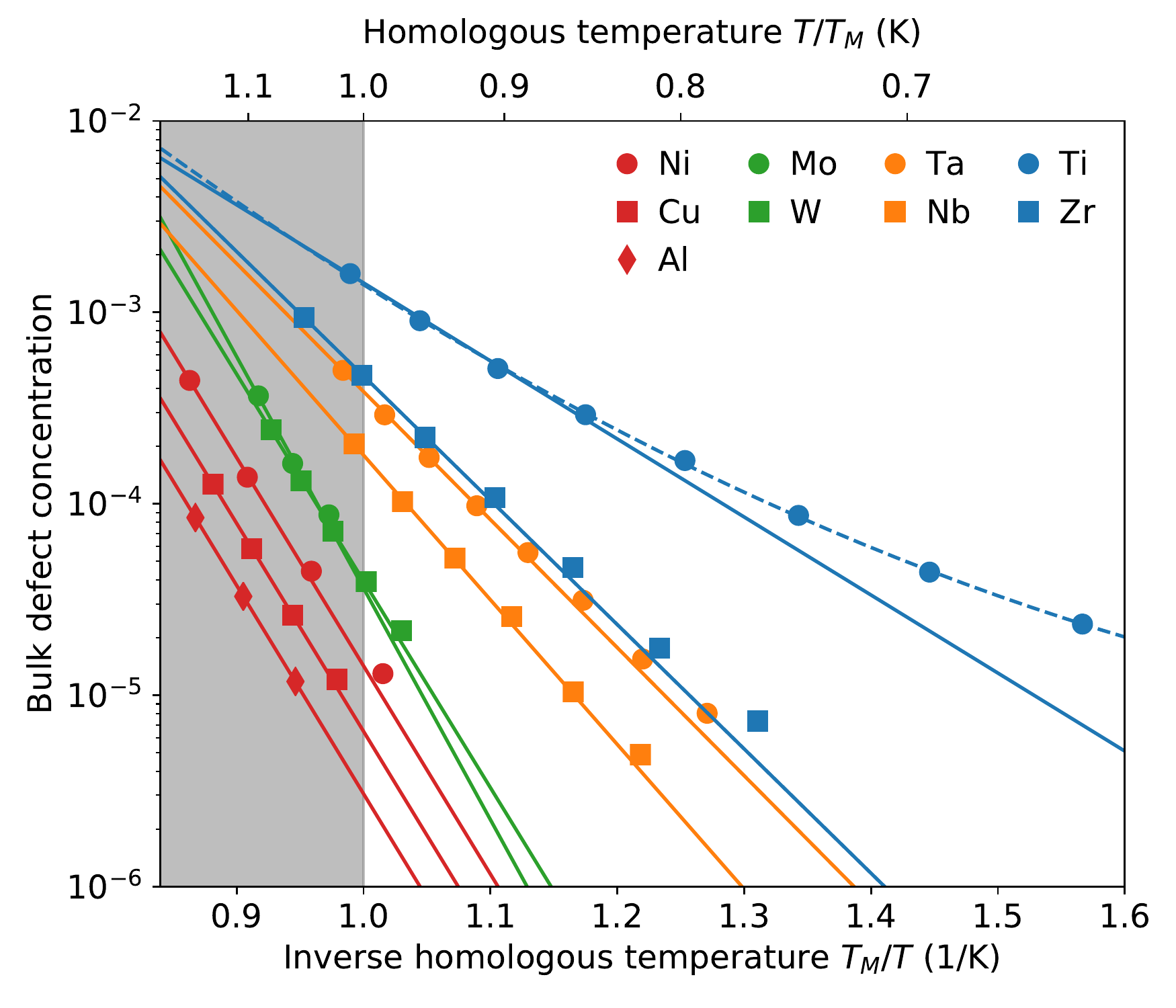}
  \caption{
    Concentration of vacancy-interstitial pairs due to spontaneous defect formation in bulk \gls{bcc} (Mo, W, Ta, Nb, Ti, Zr) and \gls{fcc} (Ni, Cu, Al) metals from \gls{md} simulations.
    Solid lines represent fits to $c\propto\exp(-E_A/k_B T)$ where $E_A$ is the activation energy for spontaneous defect formation.
    The dashed line is a higher order exponential fit to the Titanium concentrations.
    The gray shaded area marks the superheated region corresponding to temperatures above the melting point $T_m$.
    While in \gls{fcc} metals defect concentrations are very small up to the melting point, in \gls{bcc} metals defect concentration are several orders of magnitude larger and as a result are already substantial far below melting.
    Estimated 95\%\ confidence interval errors for the data points are typically substantially smaller than the symbols used in this figure.
  }
  \label{fig:defect-concentrations}
\end{figure}


\Gls{md} simulations were conducted for six \gls{bcc} and three \gls{fcc} metals to study defect formation in the bulk using different interatomic potentials \cite{HenLenTri08, MenAck07, RavGerGue13, FelParWil10, DerNguDud07a, ParFelLen12, MisFarMeh99, MenKraHao12, MisMehPap01}, sampling for up to \unit[100]{ns} and considering systems with up to 500,000 atoms.
Due to the absence of surfaces, all observed defects are the result of the spontaneous nucleation of vacancy-interstitial (Frenkel) pairs.
The concentrations of these defect pairs exhibit an exponential temperature dependence $c \propto \exp(-E_A/k_B T)$ over the entire temperature range, with the exception of the group 4 elements Ti and Zr (\fig{fig:defect-concentrations}).
The latter two metals display a deviation from exponential behavior, which we attribute to the strong anharmonicity of these materials.

Defect concentrations are particularly high in group 3 (Ta, Nb) and 4 (Ti, Zr) metals, for which values above $10^{-5}$ (the lower detection limit in our simulations) are observed down to homologous temperatures $T/T_m$ of 0.64 (Ti), 0.76-0.78 (Zr, Ta), and 0.82 (Nb).
While for vacancies the resulting concentrations are smaller than the values obtained from the defect formation energies, the situation is reversed for interstitials.
In the case of Ti and Zr, the \gls{bcc} phase only appears at high temperatures and is stabilized by strongly anharmonic lattice vibrations.
Yet high defect concentrations are also obtained for Ta and Nb, which have \gls{bcc} ground state structures at low temperatures.
For the group 6 elements (Mo, W) and even more so the \gls{fcc} elements (Al, Cu, Ni), defect concentrations are generally much smaller.
This is particularly evident at the melting point, at which the concentration of spontaneously formed Frenkel pairs varies from $10^{-3}$ for Ti to $3\times10^{-6}$ for Al, the latter value being extrapolated from higher temperatures.
The presence of these defects is actually crucial for the melting process itself, in which defects are known to play an important role \cite{SamTucYu14}.
In the case of the group 6 \gls{bcc} and \gls{fcc} elements, the relatively low defect concentrations actually enable us to superheat (on the timescale of the present simulations) the crystalline phase far into the stability range of the liquid phase.
Generally, we observe that for defect concentrations $\gtrsim\,10^{-3}$ nucleation of the liquid phase is all but inevitable.
In the case of Ti, this limit is already reached at $T/T_m\approx 1$, whereas for Al a value of $T/T_m\approx 1.35$ can be estimated by extrapolation of the available data.
The activation energies $E_A$ normalized by the melting temperature, $\alpha = E_A / k_B T_m$, exhibit similar group specific trends.
While $\alpha$ ranges from 9.4 (Ti) to 17.4 (Nb) for groups 4 (Ti, Zr) and 5 (Ta, Nb), the values for group 6 (Mo, W) as well as the \gls{fcc} metals (Al, Cu, Ni) are about 25 and larger indicating a much steeper temperature dependence.


The systematic observation of spontaneous defect formation naturally leads to the question which fundamental principles govern this behavior.
Firstly, it is instructive to explore the formation mechanism in detail.
To this end, we analyzed up to 500 trajectories for three \gls{bcc} (Ti, Ta, W) and one \gls{fcc} metal (Al) with a time resolution of \unit[10]{fs}, which allowed us to gather statistics at temperatures corresponding to a defect concentration of roughly $1.5\cdot 10^{-4}$.
In the following, we focus on the results for Ti and Al; results for other materials can be found in the Supplementary Material (Fig.~S2).

\begin{figure*}
  \raggedright
  \includegraphics[scale=0.58]{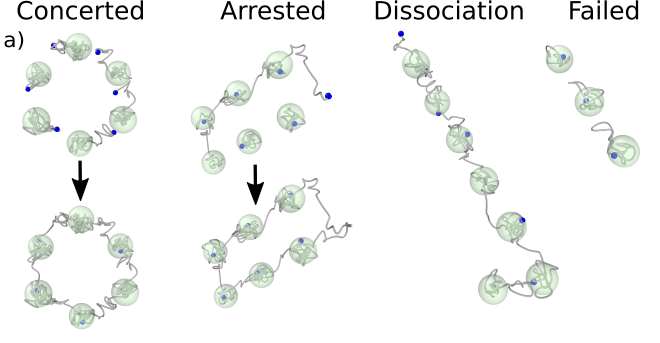}
  \includegraphics[scale=1.0]{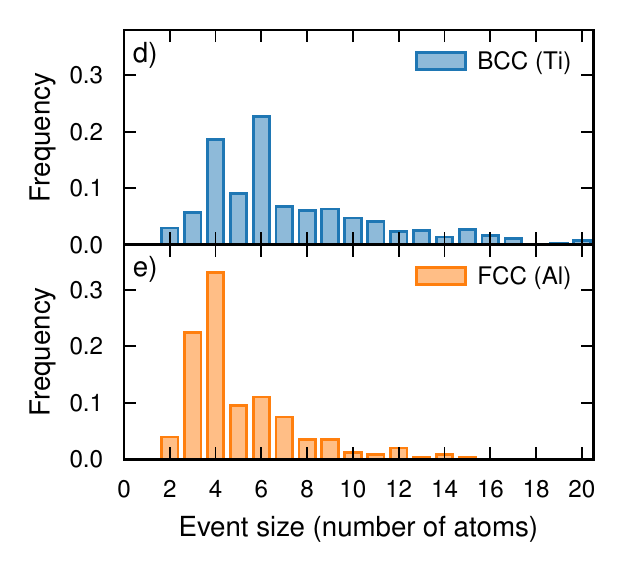}
  \includegraphics[scale=0.5]{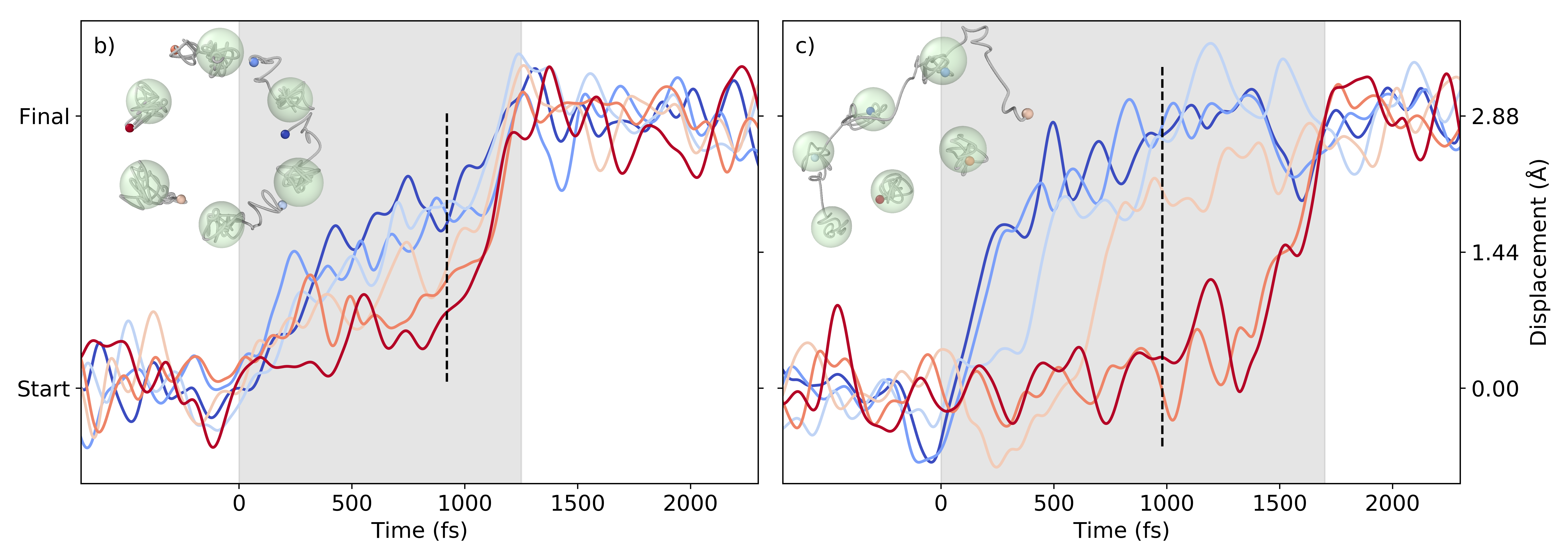}
  \caption{
    (a) Atomic scale events related to the formation of vacancy-interstitial pairs in \gls{bcc} metals.
    Small blue spheres mark atomic positions, green transparent spheres mark the initial atomic sites with the size indicating typical thermal displacements, and the gray lines indicate the atomic trajectories since the beginning of the event.
    Atomic displacements during (b) a concerted (ring-like) exchange event leading to atom exchange but no defect formation and (c) an arrested exchange event leading to the intermediate formation of an interstitial and a vacancy.
    Movies of the events are provided in the Supplementary Material.
    Frequency of events by size for (d) Ti and (e) Al.
  }
  \label{fig:events}
\end{figure*}

Both \gls{bcc} and \gls{fcc} metals exhibit ring-like exchange mechanisms involving up to approximately six atoms, after which all atoms occupy regular lattice sites again (\fig{fig:events}a,b) and Movie 1 in the Supplementary Material] \footnote{
    We note that isolated events similar to the ones described here have also been observed in ab-initio \gls{md} simulations of \gls{bcc}-Ti close to the melting point, albeit for very small cells (54 atoms) and much shorter time scales, Ref.~\onlinecite{KakHonWal17}.
}.
These events, however, do not lead to defect formation.
Alternatively, the concerted motion of a string of atoms can become arrested and even break up into substrings (\fig{fig:events}a,c and Movie 2 in the Supplementary Material], at which point the leading part of the string becomes an interstitial whereas the trailing part leaves behind a vacancy.
Subsequently, the two defects thus formed migrate independently before either recombining with each other or other defects in the system.
In the case of small ring-like exchange events, \emph{all} atoms move at the same time (\fig{fig:events}b).
By contrast in the larger events, several \emph{but not all} atoms move synchronously leading to the formation of an interstitial and a vacancy that are relatively close to each other, which can subsequently either recombine with each other by a sequence of individual defect migration events (arrested event in \fig{fig:events}a) or migrate away from each other and thus become truly independent defects (dissociation event in \fig{fig:events}a).
During their respective lifetimes these individual defects contribute to the thermodynamic equilibrium concentration of vacancies and interstitials.
Their lifetimes are determined by the time it takes to recombine with another opposing defect and are thus dependent on the overall defect concentration.
This causes the concentration of spontaneously formed defects to have a strong system size dependence and requires very large simulation cells to achieve convergence at low temperatures (see Fig.~S1 in the Supplementary Material).

The \gls{fcc} lattice structure has a higher packing density than the \gls{bcc} structure, which leads to the question whether defect formation via spontaneous atom exchanged is generally more prevalent in more open structures.
We therefore also conducted \gls{md} simulations for silicon, for which a concerted exchange event has already been proposed as a means of diffusion \cite{Pan86}.
In this case we found that 99\%\ of the observed events involved only two atoms and did \emph{not} lead to the formation of individual defects.
In the more open diamond structure two atoms can thus exchange their positions with a rather limited impact on the surrounding atoms and the probability to observe larger (string-like) events is much lower than in more densely packed structures.
This suggests that intermediate (\gls{bcc}) to high (\gls{fcc}) packing densities are crucial to obtain a larger number of events that involve several atoms.
Larger events in turn are a prerequisite to separate atoms spatially preventing immediate recombination and thus enabling the formation of independent defects.

According to a detailed analysis of the nucleation and migration events, the nucleation as well as migration of a defect pair in \gls{bcc} is predominantly the result of concerted atomic motion along close packed $\left<111\right>$ directions.
The shortest closed strings of atoms that can be constructed by concatenation of these directions involve either four or six atoms, which explains the predominance of events of these sizes in the event distribution for \gls{bcc} metals (\fig{fig:events}d).
Similarly, in \gls{fcc} systems we find defect nucleation to involve strings of atoms moving along close-packed $\left<110\right>$ directions.
The geometry of the lattice then allows for four-membered rings whereas one cannot construct equally compact six-membered rings.
Accordingly, in the event size distribution there is a pronounced peak at four.
For both \gls{bcc} and \gls{fcc} metals the distribution exhibits a heavy tail, which is the result of larger dissociation events (\fig{fig:events}a).

\begin{figure*}
  \centering
  \includegraphics{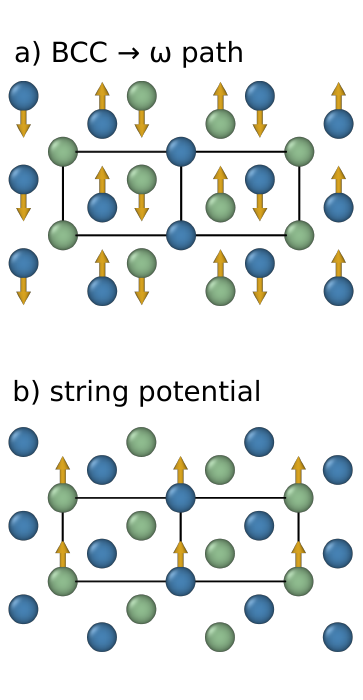}
  \includegraphics{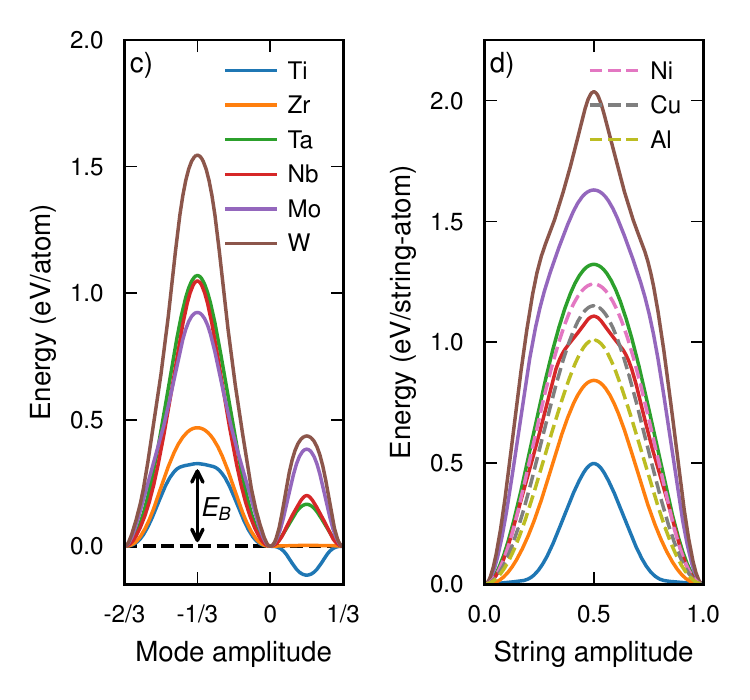}
  \includegraphics{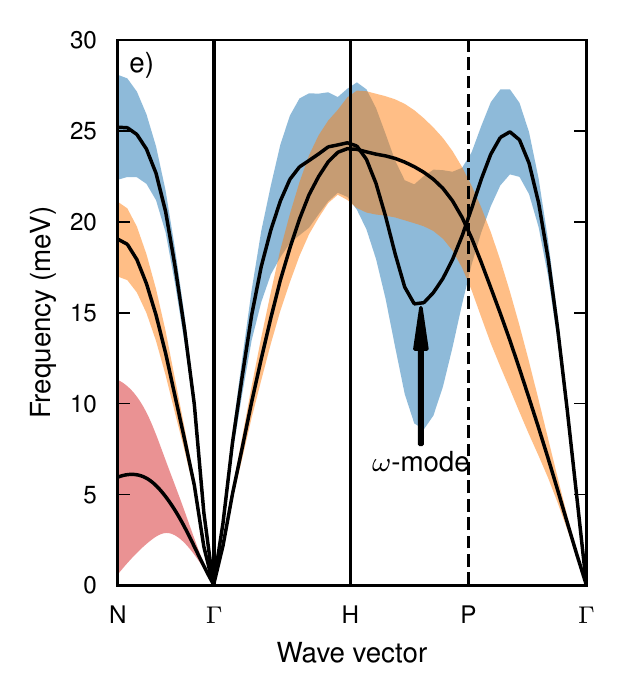}
  \caption{
    (a) Atomic displacement pattern and (c) \glspl{pes} associated with the longitudinal acoustic mode at $\vec{q}=\frac{2}{3}[1,1,1]$ corresponding to a transition path between the \gls{bcc} and $\omega$ structures.
    (b) Atomic displacement pattern and (d) \glspl{pes} associated with the string potential.
    The latter panel includes data not only for \gls{bcc} but also \gls{fcc} structures.
    The mode amplitude in (b) and the string amplitude in (d) are given in units of one repetitive unit.
    The mode and string amplitudes are normalized to the periodicity of the lattice.
    In (d) this corresponds to a displacement of the string of $\sqrt{3}/2a_0$ for \gls{bcc} and $a_0/\sqrt{2}$ for \gls{fcc} metals.
    (e) Phonon dispersion relation for \gls{bcc} Ti at \unit[1400]{K} computed from \gls{md} simulations.
    The shaded regions indicate the phonon lifetimes.
  }
  \label{fig:phonons_and_strings}
\end{figure*}
\begin{figure}
  \centering
  \includegraphics{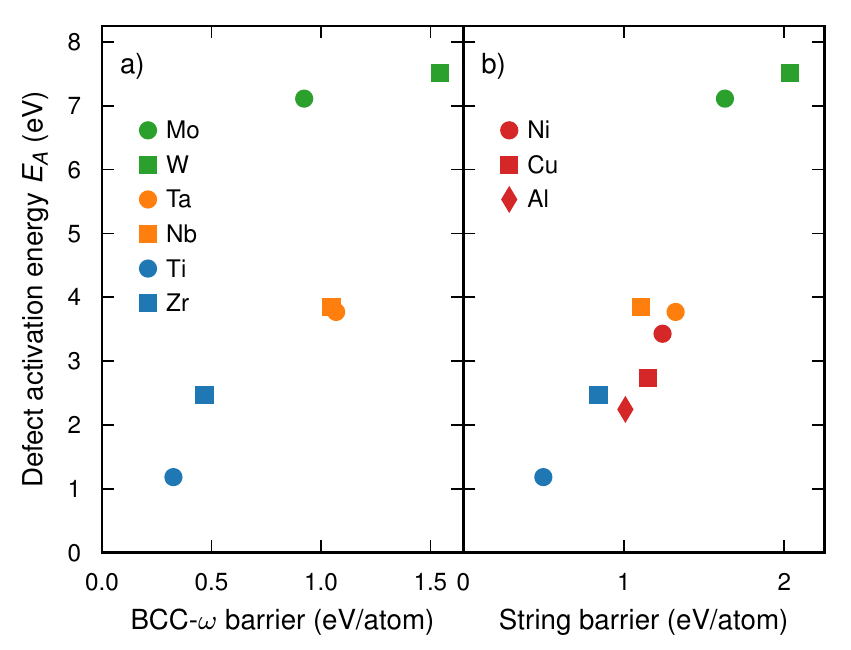}
  \caption{
    Correlation of the defect activation energy with (a) the barrier between two \gls{bcc} structures along the \bccomg{} mode (\fig{fig:phonons_and_strings}c) and (b) the string potential barrier (\fig{fig:phonons_and_strings}d).
  }
  \label{fig:barrier_correlations}
\end{figure}

The preferential motion along close-packed directions is familiar from the migration of self-interstitial atoms \cite{NguHorDud06}.
The latter exhibit crowdion characteristics in the \gls{bcc} metals considered here and their energetics can be well described using Frenkel-Kontorova models \cite{DerNguDud07a, FitNgu08}.
These models require as material specific input the so-called string or interrow potential, which describes the \gls{pes} as a function of the displacement of a string of atoms along $\left<111\right>$ (\fig{fig:phonons_and_strings}a,c).
Our calculations show that the maxima of the string potential for a set of materials correlate with the defect concentrations as well as the defect activation energies (\fig{fig:barrier_correlations}), where the latter can be extracted from the concentration vs temperature data by fitting an exponential expression (solid lines in \fig{fig:defect-concentrations}).

The parallel motion of $\left<111\right>$ strings also plays a role in the transition from the \gls{bcc} to the $\omega$ structure, which has been investigated in detail in the case of the group 4 elements \cite{TraHeiPet91, PetHeiTra91, HeiPetTra91, TriHenSri03}.
The most compact path between these structures is related to the longitudinal acoustic mode at $\vec{q}=\frac{2}{3}[1,1,1]$.
It can be represented using a 3-atom unit cell and from here on will be referred to as the \bccomg{} mode (\fig{fig:phonons_and_strings}b).
Unlike the string potential the \bccomg{} mode can be accessed experimentally and has been extensively characterized in the past \cite{TraHeiPet91, PetHeiTra91, HeiPetTra91}.
The connection between the \bccomg{} mode and defect formation is particularly apparent in the case of \gls{bcc} Ti, for which the phonon mode in question appears as a pronounced minimum along the H--P direction (\fig{fig:phonons_and_strings}e).
Following the normal coordinate of this mode reveals a strongly anharmonic \gls{pes} with a very small barrier of \unit[0.32]{eV/atom} separating periodic images of the structure (\fig{fig:phonons_and_strings}d).
Note that in these calculations we considered very large displacements beyond the range usually considered in phonon calculations, which reveal the periodic nature of the \gls{pes}.

To explore the connection between the \bccomg{} mode and the concentration of defect pairs nucleated in the bulk further, we computed the \gls{pes} also for the other \gls{bcc} metals considered in this study (\fig{fig:phonons_and_strings}d).
The maximum energy encountered along the \bccomg{} path $E_{max}^\bccomg$ does indeed correlate strongly with the activation energy $E_A$ observed for the concentration of defect pairs nucleated in the bulk (\fig{fig:barrier_correlations}b).
In the case of the group 4 metals (Ti, Zr) the \gls{pes} also features a pronounced negative region indicating the mechanical instability of the \gls{bcc} structure, which at the same time is responsible for its vibrational stabilization at high temperatures.


In conclusion, in this work we have demonstrated an unusually strong propensity of the \gls{bcc} structure for intrinsic defect formation compared to other crystal structures.
The concentration of point defects created \emph{spontaneously in the bulk} can be as large as $10^{-3}$ at the melting point and be substantial ($>10^{-5}$) down to approximately 50\%\ of the melting point (\fig{fig:defect-concentrations}).
Spontaneous defect formation proceeds via the concerted (string-like) motion of atoms, which is most likely to occur along close-packed directions (\fig{fig:events}).
In order for an initial vacancy-interstitial pair to dissociate one requires at least four to six atoms to be involved in the process, a condition that is satisfied in more densely packed crystal structures such as \gls{bcc} and \gls{fcc} (as opposed to e.g., diamond).
At the same time the barrier for atomic displacement along the close-packed direction should be sufficiently small.
As a result of this interplay, spontaneous defect formation has been found to be particular prevalent in materials that adopt the \gls{bcc} crystal structure.
The displacement of strings of atoms in the \gls{bcc} structure can be linked to the \bccomg{} phonon mode (\fig{fig:phonons_and_strings}) and one observes in fact a strong correlation between the defect formation activation energy and both string potential and \bccomg{} barrier (\fig{fig:barrier_correlations}).
This allows one to estimate the contribution of spontaneous defect formation from microscopic quantities that can be readily obtained from calculations.

\begin{acknowledgments}
This work was funded by the Knut and Alice Wallenberg Foundation and the Swedish Research Council.
Com\-puter time allocations by the Swedish National Infrastructure for Computing at NSC (Link\"oping) and C3SE (Gothenburg) are gratefully acknowledged.
\end{acknowledgments}

\section*{Appendix: Methods}

Extensive molecular dynamics (\gls{md}) simulations were carried out for a range of \gls{bcc} (Ti, Zr, Ta, Nb, Mo, W) and \gls{fcc} metals (Al, Cu, Ni) using the \textsc{lammps} code \cite{Pli95}.
Periodic boundary conditions were applied in all directions and the simulations were started from perfect (defect free) systems.
All defects observed in the simulations were thus the result of spontaneous formation in the bulk.
We sampled temperatures between 115 and 60\%\ of the melting temperature $T_m$ and conducted a careful convergence study of the results with respect to system size (see Fig.~S1 in the Supplementary Material for an example).
The atomic interactions were described using embedded atom method (Al: \cite{MisFarMeh99}, Cu: \cite{MisMehPap01}, Ni: \cite{MenKraHao12}, Ta: \cite{RavGerGue13}, Nb: \cite{FelParWil10}, Zr: \cite{MenAck07}, W:  \cite{DerNguDud07a}) and modified embedded atom method (Ti: \cite{HenLenTri08}, Mo: \cite{ParFelLen12}) potentials.
The phonon dispersion in \fig{fig:phonons_and_strings}e was obtained by \gls{md} simulations analyzed using the \textsc{dynasor} package \cite{dynasor}.
The results were analyzed using the atomic simulation environment \cite{LarMorBlo17} and \textsc{ovito} \cite{Stu10}.
Defects were identified using Voronoi analysis.

\end{document}